\documentclass{article}
\usepackage[T1]{fontenc}
\usepackage[utf8]{inputenc}
\usepackage[]{ismir} %
\usepackage{amsmath,cite,url}
\usepackage{graphicx}
\usepackage{color}

\usepackage{enumitem}
\usepackage{color}
\usepackage{subcaption}

\usepackage{amsmath,amsfonts,bm}

\def\eqref#1{equation~\ref{#1}}

\def\1{\bm{1}}

\DeclareMathAlphabet{\mathsfit}{\encodingdefault}{\sfdefault}{m}{sl}
\SetMathAlphabet{\mathsfit}{bold}{\encodingdefault}{\sfdefault}{bx}{n}

\def\gG{{\mathcal{G}}}

\usepackage{lineno}

\title{Automatic Melody Reduction via Shortest Path Finding}

\multauthor
  {Ziyu Wang$^{12}$ \hspace{1cm} Yuxuan Wu$^1$ \hspace{1cm} Roger B. Dannenberg$^3$ \hspace{1cm} Gus Xia$^1$}
  {
  $^1$ Music X Lab, MBZUAI \hspace{0.3cm} $^2$ New York University  \hspace{0.3cm} $^3$ Carnegie Mellon University \\
  {\tt\small \{ziyu.wang, yuxuan.wu, gus.xia\}@mbzuai.ac.ae, rbd@cs.cmu.edu}
  }

\def\authorname{Z. Wang, Y. Wu, R. Dannenberg, and G. Xia}

\usepackage[bookmarks=false,pdfauthor={\authorname},pdfsubject={\pdfsubject},hidelinks]{hyperref}

\begin{document}

\maketitle

\begin{abstract}
Melody reduction, as an abstract representation of musical compositions, serves not only as a tool for music analysis but also as an intermediate representation for structured music generation. Prior computational theories, such as the Generative Theory of Tonal Music, provide insightful interpretations of music, but they are not fully automatic and usually limited to the classical genre. In this paper, we propose a novel and conceptually simple computational method for melody reduction using a graph-based representation inspired by principles from computational music theories, where the reduction process is formulated as finding the shortest path. We evaluate our algorithm on pop, folk, and classical genres, and experimental results show that the algorithm produces melody reductions that are more faithful to the original melody and more musically coherent than other common melody downsampling methods. As a downstream task, we use melody reductions to generate symbolic music variations. Experiments show that our method achieves higher quality than state-of-the-art style transfer methods.\footnote{Music samples of melody reduction and variation can be found at \url{https://auto-melody-reduction.github.io/AMRA-demo/}. We release the code at \url{https://github.com/ZZWaang/melody-reduction-algo}.}

\end{abstract}

\section{Introduction}\label{sec:introduction}

Maintaining structural coherence in long-term music generation is a fundamental challenge. One approach to addressing this challenge is through hierarchical models, which rely on extracting high-level \textit{abstractions} to enable cascaded generative processes \cite{DBLP:conf/ismir/DaiJGD21, DBLP:conf/icassp/WeiXZLG22, DBLP:conf/iclr/WangMX24, DBLP:journals/corr/abs-2005-00341}. These abstractions provide a coarser-grained view of musical structure, capturing essential long-range dependencies. In existing approaches, abstractions are typically explicitly defined (e.g., chord progression or phrase labels) or learned through unsupervised methods (e.g., latent codes via an autoencoder). Yet, so far, they have not been able to capture a fundamental musical structure: the \textit{melodic flow}---how a melody evolves and resolves within a phrase---which remains too nuanced to be explicitly labeled and too challenging for unsupervised learning to reliably identify.

From a musicology perspective, melodic flow can be represented through \textit{melody reduction}, which preserves the structural essence of a melody \cite{Schenker1979FreeComposition, lerdahl1996generative}. However, most existing approaches regard melody reduction as a by-product of analysis, typically represented by hierarchical structures such as trees for further interpretation \cite{DBLP:conf/ismir/OrioR09, DBLP:conf/audio/SimonettaCOR18}. In this context, reduction is not a fixed transformation but rather a subjective and demonstrative projection of the analysis procedure. This inherent ambiguity makes melody reduction not only difficult to evaluate but also challenging to use as a practical representation \cite{DBLP:journals/ngc/TojoHH13, Groves16, DBLP:conf/ismir/Ni-HahnXYZMJR24}. In this work, we explore how melody reduction can be approximated using structural heuristics, aiming to make the concept more accessible and useful for music generation.

To this end, we propose an algorithm for automatic melody reduction. The algorithm uses the graph representation of a melody phrase and regards all possible reductions as graph paths. The intuition behind the algorithm is that if we define a cost function consistent with guiding principles underlying most reduction theories, an ideal reduction should be the path with the least cost. Specifically, we consider two principles. First, the subsequent notes in an ideal melody reduction usually reveal a simpler structure (e.g., a prolongation (unison), or a linear progression (step-wise motion) \cite{Schenker1979FreeComposition}. Second, an ideal melody reduction usually includes notes of higher significance in terms of pitch, rhythm, and harmony \cite{lerdahl1996generative, ahlback2004melody}. We define edge costs based on these principles and use the shortest-path algorithm to find the melody reduction \cite{yen1971finding}. The resulting path is subsequently post-processed into an actual melody.

We evaluate the proposed algorithm in pop, folk, and classical music genres, showing that it yields reductions that are often perceived as more faithful to the original melody and musically coherent compared to other melody downsampling methods. We also introduce variation generation as a downstream application, in which we train a melody generation model conditioned on reductions. The reductions extracted with our algorithm are shown to yield higher-quality variations compared to baselines.

\begin{figure*}[t!]

\centering

\includegraphics[width=0.79\hsize]{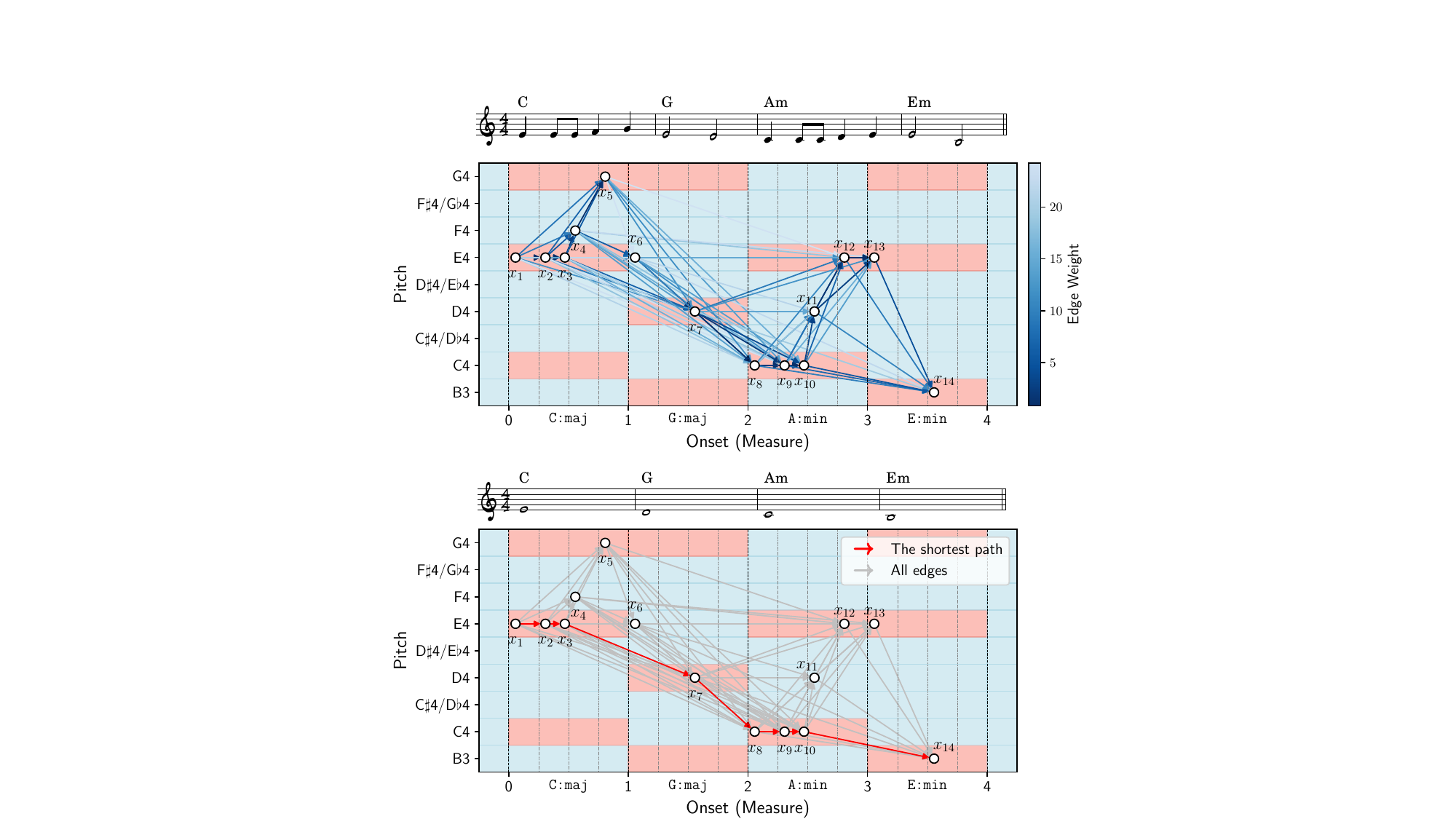}
\caption{The overview of the proposed melody reduction algorithm.}

\label{fig:diagram}
\end{figure*}

\section{Related Work}
In this section, we review three realms of related work: 1) cognitive theories about music reduction, 2) the algorithmic implementation of music theories, and 3) the importance of melody reduction in downstream applications.

In the history of cognitive music theory, a shared methodology of music analysis is to use a reduced melody to represent the abstract melodic flow \cite{Schenker1979FreeComposition, lerdahl1996generative, narmour1990analysis}. Schenkerian analysis involves a recursive reduction process to turn a music composition into the fundamental structure \cite{Schenker1979FreeComposition}; and the Generative Theory of Tonal Music (GTTM) further formalizes the grammar in Schenkerian analysis \cite{lerdahl1996generative}. These studies highlight that melody reduction is an effective representation in the cognition process, and reduction is highly related to the considerations of note connection, harmonic context, pitch importance, etc., which usually imply tension and relaxation in different music scopes. 

There are several attempts to turn these theories into algorithms. Kirlin \textit{et al.} propose a framework for automatic Schenkerian analysis \cite{KirlinU08}, and Hamanaka \textit{et al.} design an interactive software to implement GTTM using machine-learning techniques \cite{hamanaka2015gttm, hamanaka2016deepgttm, DBLP:journals/ngc/TojoHH13}. Other approaches reduce melodies recursively by assigning weights to notes \cite{DBLP:conf/ismir/OrioR09, DBLP:conf/audio/SimonettaCOR18}. However, a quality gap remains between automatic analyses and human interpretations. Moreover, the algorithms usually require score-notation level data (e.g., MusicXML format), are genre-specific, and are not open-sourced. A recent computational music analysis points out that since our music preference is hard to express in formal grammar, such formal systems tend to have a broad search space of music analyses \cite{FinkensiepR21}. This motivates us to pursue an intuitive alternative: we directly approximate melody reduction based on cognitive preference without building a formal system.

Although melody reduction is mostly studied in an analytical scope, recent advances in deep learning also show that melody reduction representation is beneficial for structured long-term music generation. Previously, melody reduction was usually implicitly modeled by surrogate features such as down-sampled melody statistics \cite{DBLP:conf/ismir/DaiJGD21}, melody contour \cite{DBLP:conf/ismir/0021WBD20}, or implicit latent representations \cite{DBLP:conf/iclr/RutteBKH23}. The recent hierarchical music generation methodology shows that using an explicitly defined melody reduction, long-term music generation can be tackled more elegantly and effectively \cite{DBLP:conf/iclr/WangMX24}. The algorithm we propose aims to establish a foundation for such future studies.

\section{Methodology}\label{sec:3:method}
In this section, we introduce the proposed melody reduction algorithm in detail. A diagram of our algorithm is shown in Figure~\ref{fig:diagram}. Section~\ref{subsec:3:repr} introduces the data attributes and the graph representation of a melody. Section~\ref{subsec:3:edge} defines the edge types of the graph, and Section~\ref{subsec:3:cost} defines the edge cost. Finally, we discuss the melody reduction post-processing operations in Section~\ref{subsec:3:post}.

\subsection{Graph Representation of a Melody}\label{subsec:3:repr}
The input to the algorithm is a sequence of notes, denoted by $x_1, ..., x_N$, and an underlying chord progression, denoted by $c_1, ..., c_K$. A melody can be represented by a directed graph $\gG(V, E)$, where the melody notes are regarded as graph nodes $V:=\{x_i\}_{i=1}^N$, and temporal relations of notes can be represented by edges $E:=\{x_i \to x_{i + 1}\}_{i=1}^{N-1}$.

We consider the onset, pitch, and duration attributes of a note $x_i$. These are denoted by $\text{Onset}(x_i), \text{Pitch}(x_i)$ and $\text{Dur}(x_i)$, respectively. The onset and duration should be quantized by beat locations, and the pitch is represented by MIDI note numbers from 0 to 127. Additionally, a chord is represented by a 12-d binary chroma vector, i.e., $c_i \in \{0, 1\}^{12}$, and we define $\text{Chord}(x_i) \in \{c_1, ..., c_K\}$ to indicate chord membership of the note $x_i$. In our algorithm, we heuristically detect anticipation-like cases (a type of non-chord tone) and regard these notes as belonging to the next chord.

\subsection{Edge Definition}\label{subsec:3:edge}
In the proposed algorithm, we regard both the original melody and reduced melodies as paths from $x_1$ to $x_N$. The original melody uses edges in $E$, whereas a reduction uses shortcut edges. To this end, we define an \textit{augmented} edge set $E^*:=\{x_i \to x_{j}|i < j\}$, representing all causal edges. If an edge $x_i\to x_j$ is selected in the reduction process, it means the melodic movement from $x_i \to x_j$ is more significant than all other movements $x_{i'} \to x_{j'}$ inside the time range, i.e., $i \leq i' < j' \leq j$ and $(i, j) \neq (i', j')$.

We categorize an edge $x_i\to x_j \in E^*$ into six categories. The first three categories are the most fundamental, which correspond to three main ways of melody reduction in Schenkerian analysis: prolongation, linear progression, and arpeggiation \cite{schenker}. Note that the edges are strictly defined below, and we only borrow the terms for implication:

\begin{itemize}
    \item \textbf{Prolongational Edge (PE)}: $x_j$ prolongs $x_i$ with the same pitch. For example, a PE can potentially remove a neighbor tone. Mathematically, a PE satisfies $\text{Pitch}(x_i) = \text{Pitch}(x_j)$ and $\text{Onset}(x_j) -  \text{Onset}(x_i) < D$.
    
    \item \textbf{Linear Edge (LE)}: the interval between $x_i$ and $x_j$ is a second. For example, an LE can potentially mark a significant melodic movement. Mathematically, an LE satisfies $|\text{Pitch}(x_i) - \text{Pitch}(x_j)| \in \{1, 2\}$ and $\text{Onset}(x_j) - \text{Onset}(x_i) < D$.
    
    \item \textbf{Arpeggiation Edge (AE)}: the interval between $x_i$ and $x_j$ is larger than a (compound) second and $x_i$ and $x_j$ are within the same chord. For example, an AE can potentially mark an elaboration of harmony. Mathematically, an AE satisfies $|\text{PitchClass}(x_i) - \text{PitchClass}(x_j)| \in \{3, 4, 5, 6, 7, 8, 9\}$ and $\text{Chord}(x_i)=\text{Chord}(x_j)$.
\end{itemize}

In some melody compositions, pitches that span an octave are also regarded as a smooth connection. In Schenkerian analysis, this is explained by \textit{imaginary continuo}---although the two tones span an octave in the current realization, they are close in other imaginary realizations. We define two types of imaginary edges accordingly:

\begin{itemize}
    \item \textbf{Imaginary Prolongational Edge (IPE)}: $x_j$ prolongs $x_i$ with the same pitch class. Mathematically, an IPE (is not a PE) and satisfies $\text{PitchClass}(x_i) = \text{PitchClass}(x_j)$ and $\text{Onset}(x_j) -  \text{Onset}(x_i) < D$.
    
    \item \textbf{Imaginary Linear Edge (ILE)}: the interval between $x_i$ and $x_j$ (or its inversion) is a compound second. Mathematically, an ILE (is not an LE) and satisfies $|\text{PitchClass}(x_i) - \text{PitchClass}(x_j)| \in \{1, 2, 10, 11\}$ and $\text{Onset}(x_j) -  \text{Onset}(x_i) < D$.
\end{itemize}

Finally, all the rest of the edges in $E^*$ belong to the final category. This is to ensure the graph is connected so that there must exist at least one path from $x_1$ to $x_N$:

\begin{itemize}    
    \item \textbf{Unclassified Edges (UE)}: the rest of the edges. 
\end{itemize}

In the above definition, $\text{Pitch}(\cdot)$, $\text{Onset}(\cdot)$, and $\text{Chord}(\cdot)$ are previously defined in Section~\ref{subsec:3:repr}. $\text{PitchClass(x)}:=\text{Pitch}(x) \mod 12$ and $\text{Chord}(x_i)=\text{Chord}(x_j)$ if and only if $x_i$ and $x_j$ are within the interval of a single chord. In our experiment, the temporal threshold $D$ is set to 2 measures.

\subsection{Edge Cost Definition}\label{subsec:3:cost}

We define the edge cost function of $x_i\to x_j$ so that a more significant edge will have a smaller cost. The edge cost function considers three aspects: 1) the function of different edge types, 2) the temporal distance of an edge, and 3) the note importance.\footnote{Currently, the costs are empirically specified based on domain knowledge and preliminary analysis. Estimating them from data is left for future work.}

First, we define \textit{tonal cost}, denoted by $c_\text{tonal}(x_i \to x_j)$, prioritizing prolongational and linear edges in the melody reduction. Formally, 

\begin{equation}
c_{\text{tonal}}(x_i \to x_j):= \begin{cases}
0.1\text{,} &\text{if } x_i \to x_j \text{ is a PE,} \\
0.3\text{,} &\text{if } x_i \to x_j \text{ is an LE,} \\
1.5\text{,} &\text{if } x_i \to x_j \text{ is an AE,} \\
1.0\text{,} &\text{if } x_i \to x_j \text{ is an IPE,} \\
1.3\text{,} &\text{if } x_i \to x_j \text{ is an ILE,} \\
3.0\text{,} &\text{if } x_i \to x_j \text{ is a UE.} \\
\end{cases}
\end{equation}

Then, we define the \textit{temporal cost}, denoted by $c_{\text{temp}}(x_i \to x_j)$, to measure the distance from index $i$ to index $j$. We set the hyperparameter $\eta=1.6$ to achieve an ideal degree of reduction. A larger $\eta$ results in too little reduction and a smaller $\eta$ makes the reduction too coarse:
\begin{equation}
c_{\text{temp}}(x_i \to x_j) := (j - i)^\eta\text{.}
\end{equation}

Besides the two cost functions on edges, we introduce a \textit{note importance factor}, denoted by $\alpha(x_i)$, to ensure structurally important notes are more likely to be selected. Particularly, $\alpha(x_i)$ is a product of four terms: 
\begin{equation}
    \alpha(x) := \alpha_\text{p}(x)\alpha_\text{o}(x)\alpha_\text{d}(x)\alpha_\text{h}(x)\text{,}
\end{equation}
where $\alpha_\text{p}(x_i)$ denotes \textit{pitch importance}, $\alpha_\text{o}(x_i)$ denotes \textit{onset importance}, $\alpha_\text{d}(x_i)$ denotes \textit{duration importance}, and $\alpha_\text{h}(x_i)$ denotes \textit{harmony importance}. 

\begin{enumerate}
    \item  \textbf{Pitch Importance.} Higher and lower pitches are usually more significant in a melody and should be given a smaller weight factor:
\begin{equation}
\alpha_\text{p}(x_i) := 0.1 \times \Bigl(0.5 - \frac{|\text{Pitch}(x_i) - p_{\text{mid}}|}{p_{\text{max}} - p_{\text{mid}}}  \Bigr) + 1\text{,}    
\end{equation}
where $p_{\text{max}}$ and $p_{\text{min}}$ are maximum and minimum pitch values and $p_\text{mid}=(p_{\text{max}} + p_{\text{min}}) / 2$. 

\item \textbf{Onset Importance.} The notes having higher metrical importance should be given a smaller weight factor:
\begin{equation}
\alpha_\text{o}(x_i) := \begin{cases}
0.85\text{,} &\text{Onset}(x_i) \in \text{DB,} \\
0.95\text{,} &\text{Onset}(x_i) \in \text{B,} \\
1.05\text{,} &\text{Onset}(x_i) \in\text{B/2,} \\
1.15\text{,} &\text{Onset}(x_i)\in \text{B/4\text{.}}  \\
\end{cases}
\end{equation}
Here, DB, B, B/2, and B/4 represent downbeat, beat, eighth-note, and sixteenth-note positions, respectively (if under the 4/4 time signature).
\item \textbf{Duration Importance.} Longer notes should be given a smaller weight factor:
\begin{equation}
\alpha_\text{d}(x_i) := \begin{cases}
0.85\text{,} &\text{Dur}(x_i) \geq \text{half note,} \\
0.95\text{,} &\text{Dur}(x_i) \geq \text{quarter note,} \\
1.05\text{,} &\text{Dur}(x_i) \geq \text{8$^\text{th}$ note,} \\
1.15\text{,} &\text{Dur}(x_i) \geq \text{16$^\text{th}$ note.}  \\
\end{cases}
\end{equation}

\item \textbf{Harmony Importance}. A chord tone should be given a smaller weight factor than non-chord tones. 
\begin{equation}
    \alpha_\text{h}(x_i) := \begin{cases}
0.85\text{,} &x_i \text{ is a chord tone,} \\
1.15\text{,} &\text{ otherwise.} \\
\end{cases}
\end{equation}
Here $x_i$ is a chord tone strictly means $\text{PitchClass}(x_i)$ is in $\text{Chord}(x_i)$. So, an anticipation is regarded as a chord tone to the next chord (see Section~\ref{subsec:3:repr}).
\end{enumerate}

Finally, the total edge cost is defined as a summation of tonal and temporal cost, modulated by the note importance factor:
\begin{equation}
c(x_i \to x_j)= \alpha(x_j)[c_\text{temp}(x_i \to x_j) + c_\text{tonal}(x_i \to x_j)]\text{.}
\end{equation}
Thus, the melody reduction can be achieved by running a shortest-path algorithm to find the shortest path from $x_1$ to $x_N$.

\begin{figure}[t!]

\centering
\includegraphics[width=\hsize]{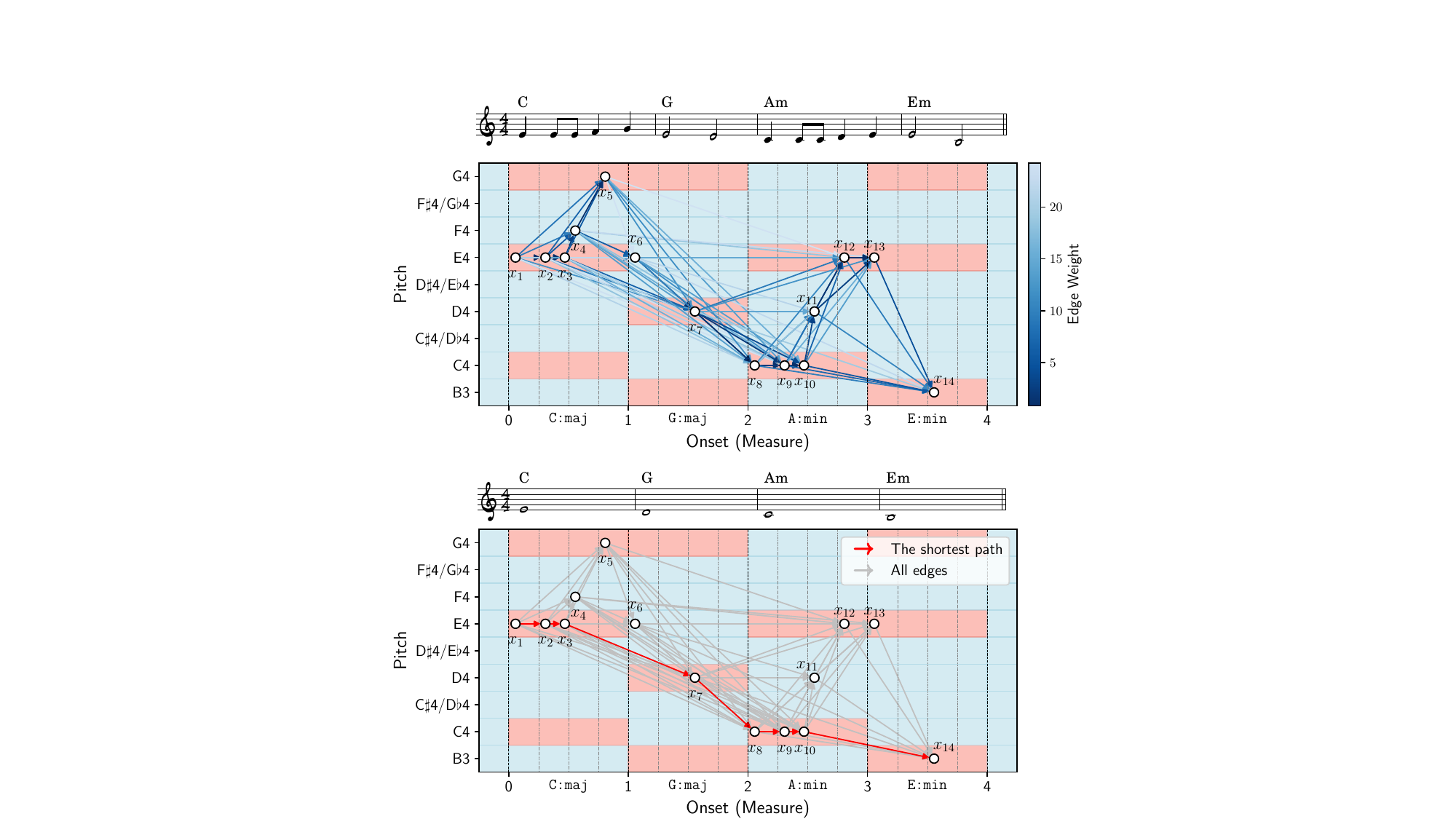}
\caption{An illustration of post-processing operations.}
\label{fig:postprocess}
\end{figure}

\begin{figure*}[t!]
    \centering
    \subfloat[Pop Genre]{\includegraphics[width=0.32\textwidth]{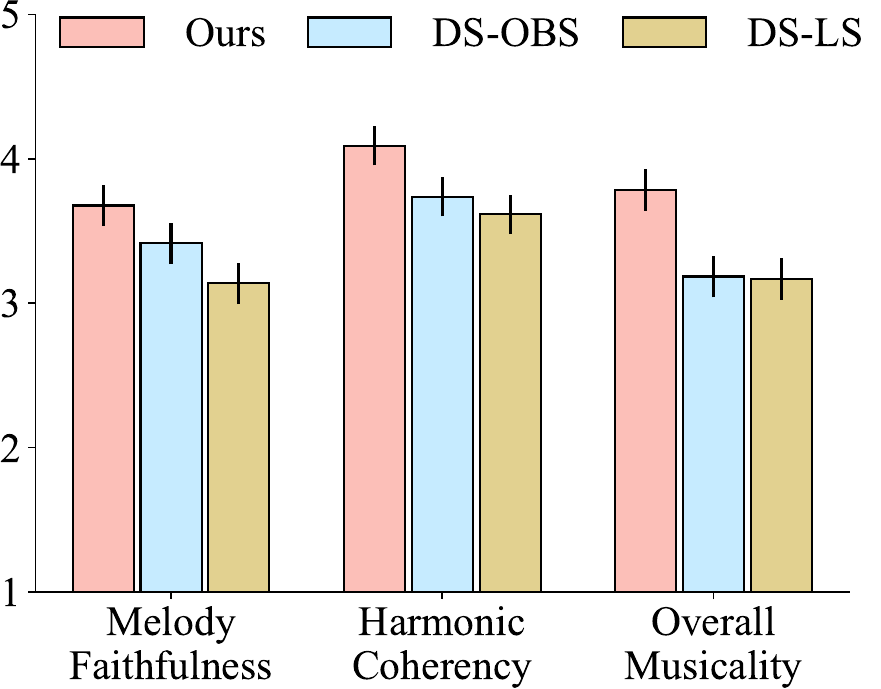}}
    \hfill
    \subfloat[Folk Genre]{\includegraphics[width=0.32\textwidth]{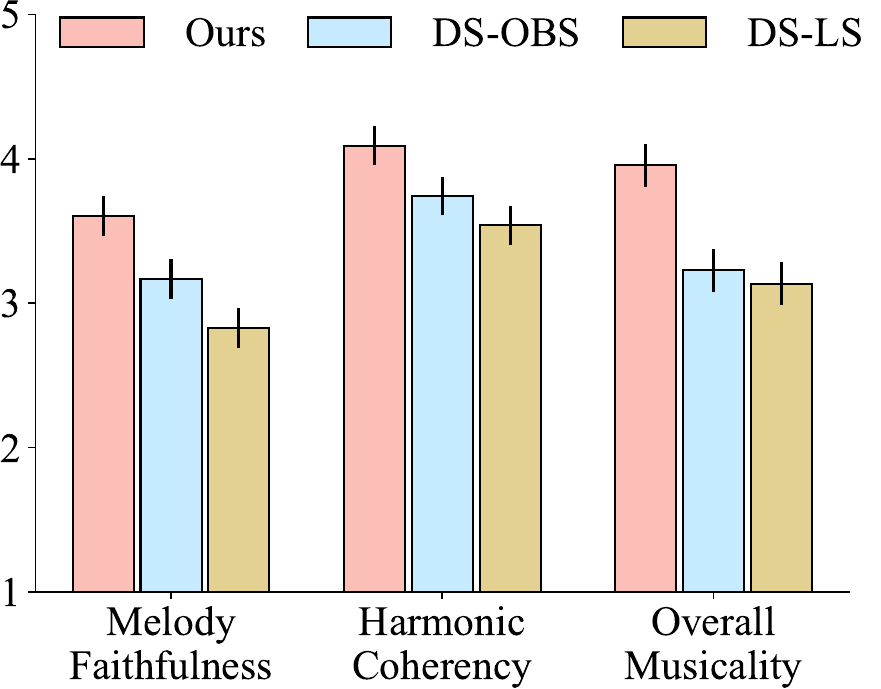}}
    \hfill
    \subfloat[Classical Genre]{\includegraphics[width=0.32\textwidth]{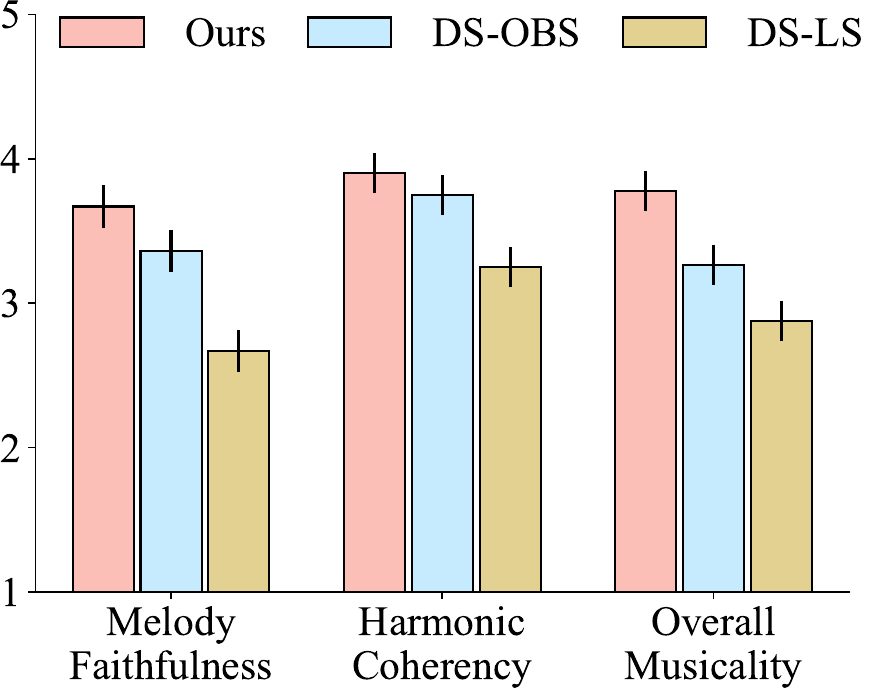}}
    \caption{Subjective evaluation results of melody reduction quality across three genres.}
    \label{fig:subjective_red}
\end{figure*}

\subsection{Post-Processing}\label{subsec:3:post}
After we find the shortest path, we use a rule-based post-processing method to arrange the selected notes in the path to melody reduction. The maximum resolution of the reduction is a quarter note, in the style of a fifth-species counterpoint \cite{fux}.

Figure~\ref{fig:postprocess} shows the detailed procedure. First, the nodes in the shortest path are allocated into \textit{chord bins}, with each bin corresponding to a distinct chord. In each chord bin, the notes are given a fixed rhythm template (see the table at the bottom of Figure~\ref{fig:postprocess}), ensuring the notes within a bin collectively span the entire duration of their associated chord. In this process, notes linked by a prolongational edge are merged into a single note. If the number of nodes in a chord bin exceeds the length of the chord, a random selection of notes will be omitted. Finally, the prolongational edges between two chords are marked with suspension. Note that notes serving as anticipations are allocated to the bin of the subsequent chord. 

\section{Experiments}\label{sec:4:expr}
In Section~\ref{subsec:4:subeval-tra}, we evaluate the proposed algorithm through a subjective listening test. In Section~\ref{subsec:4:red-sample}, we show and evaluate a melody reduction example as a case study. Finally, we evaluate the effectiveness of melody reduction in downstream music generation tasks in Section~\ref{subsec:4:gen}.

\subsection{Subjective Evaluation of Melody Reduction}\label{subsec:4:subeval-tra}

Unlike tasks with clear ground truths, melody reduction is inherently subjective and style-dependent. Existing theories, such as GTTM or Schenkerian analysis, provide interpretive hierarchies rather than prescriptive outcomes \cite{schenker, lerdahl1996generative}. Finding reduction typically involves pruning a tree at variable depths, often informed by human judgment. Moreover, such theories are primarily suited to classical music.

Given these challenges, we adopt a subjective listening test to better capture the perceptual and musical quality of melody reductions. We cover three music genres: pop, folk, and classical. We sample melodies from the POP909 dataset \cite{pop909-ismir2020}, the Nottingham dataset \cite{foxley2011nottingham}, and the GTTM database \cite{GTTMDatabase} for the pop, folk, and classical genres, respectively. We compare with two representative baselines commonly used for melody reduction as feature extraction in music generation:
\begin{itemize}[itemsep=2pt, topsep=2pt, parsep=0pt]
    \item \textbf{Downsampling on Observations} (\textit{DS-OBS}): From a statistical perspective, the melody is downsampled to a sequence of half notes, each representing the most common pitch in the 2-beat music segment \cite{DBLP:conf/ismir/DaiJGD21}. 
    \item \textbf{Downsampling in Latent Space} (\textit{DS-LS}): EC$^{2}$-VAE \cite{ec2vae} learns disentangled latent representations of the pitch contour and rhythmic pattern of 2-measure music segments as $z_{\text{p}}$ and $z_{\text{r}}$, respectively, which enables downsampling in the latent space of rhythm patterns. Specifically, we encode the pitch contour $z_{\text{p}}$ of data and decode it together with a downsampled rhythm $z_{\text{r}}$ to get the melody reduction for every 2-measure segment.
\end{itemize}

For the subjective test, we randomly select four 8-measure melodies from each genre. Each participant listens to at least 3 groups of melody reductions for each genre. In each group, participants are presented with the original melody first, followed by the melody reductions generated by the proposed algorithm, downsampling, and latent representation recombination in a randomized order. Participants are asked to rate the quality of the melody reduction on a 5-point Likert scale, where 1 indicates the worst quality and 5 indicates the best, in terms of three criteria:
(1) \textit{Melody Faithfulness}: how well the melody reduction preserves the original music information. 
(2) \textit{Harmonic Coherency}: how well the melody reduction fits the underlying chord progression.
(3) \textit{Overall Musicality}: the overall music quality of the melody reduction.

A total of 45 subjects (26 females \& 19 males) participated in the survey, in which over 70\% have a music education experience of at least 2 years.
The results are reported in Figure~\ref{fig:subjective_red}, where the heights of bars represent means of the ratings and the error bars represent the standard error computed by within-subject ANOVA \cite{scheffe1999analysis}. The proposed algorithm is significantly preferred over both baselines in all genres and criteria ($p < 0.05$), except for melody faithfulness in the pop genre, where the difference shows a positive but marginal trend ($p < 0.075$).

\subsection{A Case Study of Melody Reduction}\label{subsec:4:red-sample}

\begin{figure}[t]
    \centering
    \includegraphics[width=1\hsize]{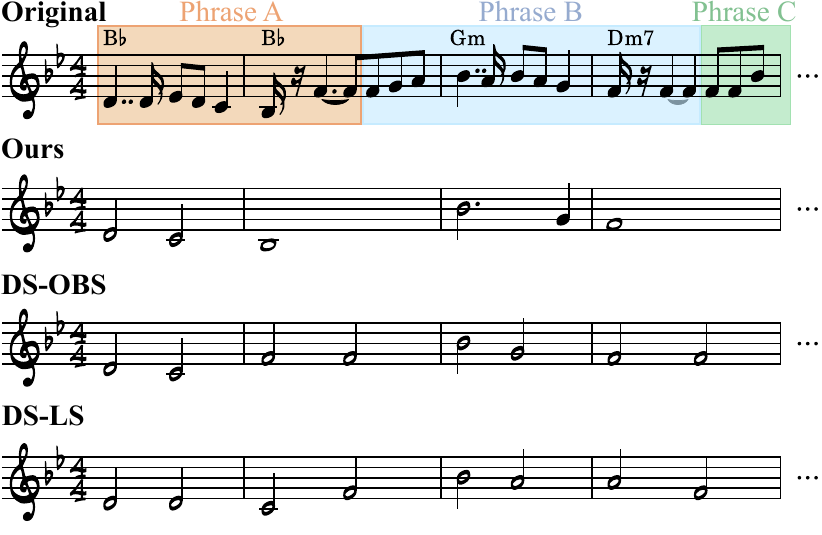}
    \caption{Comparison of the original melody, melody reductions from the proposed method, and the baselines. We highlight the phrases in the top row.
    }
    \label{fig:example_red}
\end{figure}

We provide a case analysis of melody reduction comparing the proposed algorithm and baselines introduced in Section~\ref{subsec:4:subeval-tra}, as shown in Figure~\ref{fig:example_red}. The original melody is shown in the top row, followed by the three melody reductions generated by the proposed algorithm and baselines.

\begin{figure*}[t!]
    \centering
    \includegraphics[width=\textwidth]{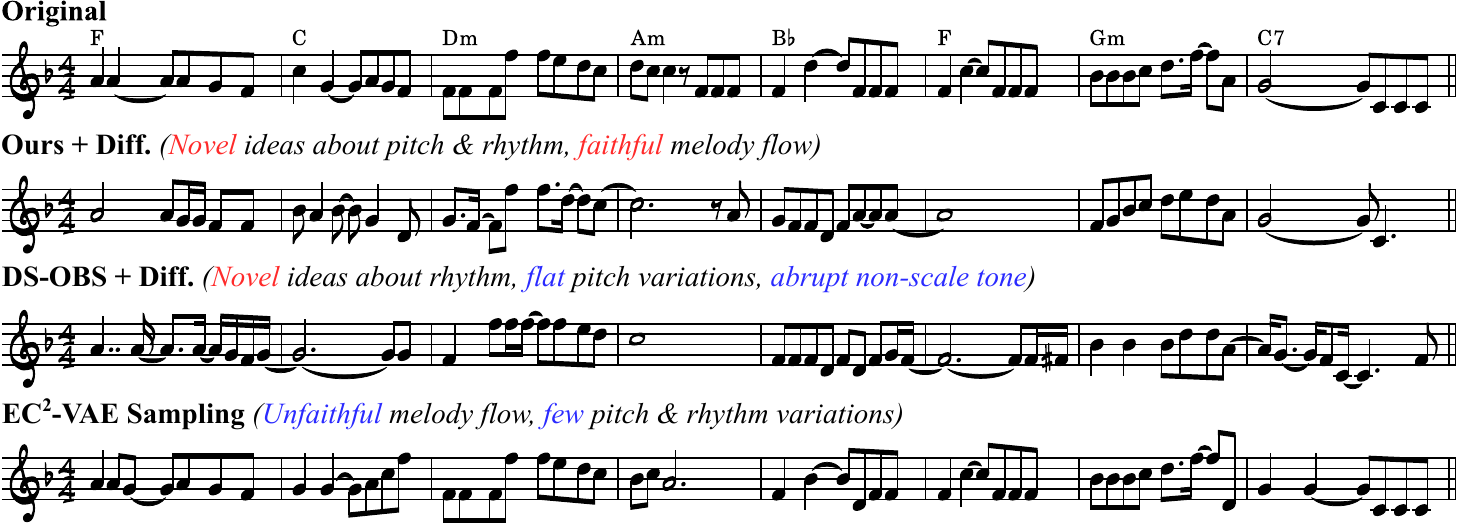}
    \caption{Comparison of variations of an example melody. Here \textbf{Diff.} denotes the conditional diffusion model trained to generate melody variations from melody reductions. Positive comments are highlighted in red, negatives in blue.} 
    \vspace{-0.2cm}
    \label{fig:example_gen}
\end{figure*}

Both the proposed method and DS-OBS can mostly capture the correct melody flow, such as important passing tones like C4 in the first measure. In subtle situations such as the second measure, where Phrase A ends its downward music flow with the downbeat chord tone B$\flat$3 and lingers at F4 until the transition to Phrase B, DS-OBS fails to preserve B$\flat$3, as it is overshadowed by the long duration of F4. In contrast, the proposed algorithm successfully captures B$\flat$3 by paying attention to its harmonic and rhythmic importance, as well as the imaginary prolongational edge between B$\flat$3 and B$\flat$4 in the third measure. DS-LS only captures the pitch contour but introduces several unwanted non-chord tones. This example demonstrates the effectiveness of the proposed algorithm for melody reduction.

\subsection{Downstream Task: Generating Melody Variations}\label{subsec:4:gen}

We believe melody reduction can serve as a useful representation of structural information in downstream tasks. In this section, we demonstrate one such application in a melody variation generation task. The task uses the reduction of a melody as input, and outputs variations faithful to the original melody. While we do not claim a strong causal link between reduction quality and generation quality, our intuition is that an accurate reduction better reflects the underlying melodic and harmonic context, which in turn supports more coherent and musically grounded generation.

To this end, we train a diffusion model to generate melody variations from melody reductions provided by the proposed algorithm. We use a similar model design and training settings as the leadsheet generation model in \cite{DBLP:conf/iclr/WangMX24} and train the model on the POP909 dataset. Similarly, we train the model using melody reductions by DS-OBS. We also generate melody variations by sampling in the latent space of $z_{\text{p}}$ and $z_{\text{r}}$ of EC$^{2}$-VAE for comparison. For all methods, we randomly sample four outputs per input and select the most representative one for use in listening tests.

Figure~\ref{fig:example_gen} shows a group of melody variation examples. It can be seen that the variation model trained with the proposed melody reductions not only maintains the original melody flow but also introduces novel ideas in pitch and rhythm. The model trained with DS-OBS also preserves the pitch contour, but tends to have flat pitch variations. The variation generated by sampling from the latent space of EC$^{2}$-VAE changes the original melody flow in an unwanted way, and does not introduce rich variations.

\begin{figure}[t!]
    \centering
    \includegraphics[width=0.99\hsize]{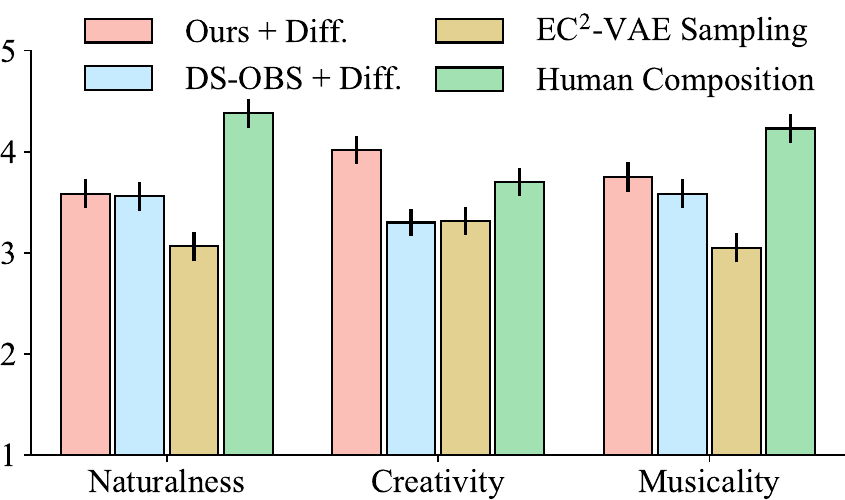}
    \caption{Subjective results of melody variations.}
    \label{fig:subjective_gen}
\end{figure}

We evaluate the melody variations on the test set of POP909 using a subjective listening test with the same participants as in Section~\ref{subsec:4:subeval-tra}. Each participant listens to at least three groups of melody variations, where participants are first presented with the original melody, followed by the three variations in random order. Participants are asked to rate the quality of melody variations and the original melody by human composers in three criteria: \textit{Naturalness}, \textit{Creativity}, and \textit{Musicality} \cite{chu2022empirical}. The results are reported in Figure~\ref{fig:subjective_gen}, with the same computation as in Section~\ref{subsec:4:subeval-tra}, including mean ratings and statistical significance tests. Our method is consistently preferred over two baselines in terms of creativity and overall musicality ($p < 0.05$), and remains competitive in naturalness.

\section{Conclusion}
To sum up, this work proposed a novel and useful algorithm for melody reduction, filling the gap between the need to capture melody flow for long-term and hierarchical music generation and the lack of all-genre off-the-shelf tools for melody reduction. The proposed algorithm finds the optimal melody reduction by finding the shortest path in a graph representation of the melody, considering the tonal, temporal, and note importance factors. Subjective experiments demonstrated that our method outperforms baselines in a variety of musical styles. We also demonstrated the effectiveness of the melody reduction algorithm in melody variation generation through subjective evaluation. In the future, we plan to tackle reduction that captures latent polyphony and hierarchical structure, and explore the application of the proposed algorithm in a broader range of music generation tasks. While the current algorithm contains ad-hoc parameters, future work could also explore learning these directly from data.

\bibliography{ISMIRtemplate}

\end{document}